\documentclass[12pt,a4paper,english]{article}
\usepackage{mathptmx,helvet,courier,float,textcomp,setspace,geometry,array,babel,chemformula}
\usepackage{graphics,fancybox,fancyhdr,wrapfig,soul}
\usepackage[T1]{fontenc}
\usepackage[latin9]{inputenc}
\usepackage[final]{pdfpages}
\usepackage{fix-cm,amsmath}
\usepackage{rotating}
\usepackage{amssymb}
\usepackage{xcolor}
\usepackage{cite}

\makeatletter

\newcommand{\angstrom}{\text{\normalfont\AA}}

\newcommand{\address}[1]{\vspace{-2.8em}\begin{center} #1 \end{center}\begin{large}\end{large}}
\begin{document}
\pagestyle{plain}
\pagenumbering{arabic}

\title{\textbf{\large Suppressed Defect Emission and Enhanced Blue Luminescence of Zn-Doped CuBr Quantum Dots synthesized in Room-Temperature Synthesis}}
\author{s}

\author{\normalsize Aryamol Stephen$^{a}$, Ruchi Kumari$^{b,}$, Saparja Roy$^{b,}$, P. Jayaram$^{a}$, A. Biju {$^{a,*}$} P. M. Sarun{$^{b,}$}\thanks{Corresponding Author's contact: \\ \indent\hspace{5mm} sarun@iitism.ac.in, Tel: +91--326--2235887;  bijuanchal@gmail.com, Tel:+91--944--752--0841}}

\date{\small}

\maketitle
\address{\footnotesize $^{a}$ Materials and Condensed Matter Physics Laboratory, Department of Physics, MES Ponnani College, Ponnani, University of Calicut, Malappuram, Kerala, 679586, India.}
\vspace*{0.5cm}
\address{\footnotesize $^{b}$ Functional Ceramics Laboratory, Department of Physics, Indian Institute of Technology (Indian School of Mines), Dhanbad 826004, India.}
\doublespacing
\begin{abstract}
\hspace{5mm}

Deep-blue-emitting CuBr quantum dots (QDs) hold promise as nontoxic and heavy-metal-free optoelectronic materials, yet their efficiency is limited by intrinsic defects. We report Zn doping as an effective strategy to mitigate defect-mediated recombination in CuBr QDs synthesized via an eco-friendly supersaturated recrystallization method under ambient conditions. XRD shows lattice contraction, Rietveld refinement confirms substitutional doping, and XPS verifies Cu$^+$/ Zn$^{2+}$ states without secondary phases. HR-TEM analysis reveals uniform 3.46 $\pm$ 0.12 nm Zn:CuBr QDs with reduced bandgap, after doping, indicating weaker quantum confinement. The dominant emission peaks appear at 406 nm ($Z_f$) and 431 nm ($Z_{1,2}$) under 365 nm excitation. Narrows defect FWHM, elevated band-edge emission, and extends carrier lifetimes, achieving 96.4 \% color purity with CIE coordinates (0.1520, 0.0395). These findings demonstrate that Zn doping is a simple, scalable, and environmentally benign strategy for engineering efficient blue-emitting CuBr QDs.


\end{abstract}
\newpage

\section*{\normalsize Introduction}
\hspace{5mm} 
High-efficiency, narrow-band, deep-blue emitters are essential for advancing optoelectronic technologies, including solid-state lighting, high-density optical data storage, and high-resolution displays \cite{renj2025deep, sree2025rece, tang20234dop, huat2024deep}.
At present, Cd-and Pb-based semiconductors deliver excellent optical performance and highly saturated blue emission, but their toxicity raises serious environmental and health concerns, limiting their widespread use \cite{heho2025envi,leew2020synt, lixi2020cdse}. As a result, alternative material systems have been extensively explored. Among them,  III-nitride semiconductors, particularly InGaN alloys, dominate commercial blue emitters in high-brightness LEDs and laser diodes, while II-VI compounds (ZnO, ZnS, ZnSe, ZnTe) are widely employed in blue-emitting flat-panel displays, thin-film electroluminescent devices, and lasing applications. However, these materials face significant challenges, including lattice-mismatch-induced strain in nitrides, reduced optical gain caused by internal electric fields, and difficulties in achieving stable \textit{p}-type conductivity in ZnO \cite {yili2000thep, cowl2010elec}. The increased demand for deep blue-emitting materials has driven extensive research into efficient blue light-emitting systems, among which colloidal quantum dots (QDs) have emerged as particularly promising candidates. 

Owing to their better photoluminescence, narrow emission bandwidths, excellent color purity, robust photostability, size-dependent bandgap tunability, and compatibility with solution processing, blue light-emitting QDs offer significant advantages over conventional emitters \cite {liru2025mult,heho2025envi,zhan2023coll}. These attributes make them highly attractive for applications in light-emitting diodes (LEDs), lasers, sensors, photovoltaics, photocatalysis, biomedicine, and emerging quantum technologies \cite {piet2016spec, wang2015brig, zhou2015towa, kaga2020coll}. However, achieving efficient deep-blue emission in QDs remains challenging. Short-wavelength emission requires ultrasmall QD cores, which increase the surface-to-volume ratio and generate abundant trap states. These defects drive nonradiative recombination and self-absorption, severely compromising photoluminescence efficiency \cite{kirk2018find, kimh2019emis}. Surface passivation, often achieved through controlled doping, is therefore essential. Yet, synthesizing uniformly small, doped blue-emitting QDs remains difficult and typically involves complex reaction conditions \cite{makk2018fron}.

Among emerging heavy-metal-free systems, copper(I) halides (CuX, X = Br, Cl, I) have attracted increasing attention as wide-bandgap I-VII semiconductors suitable for short-wavelength emission. In particular, CuBr exhibits a direct bandgap (3.1 eV) \cite{cowl2010elec}, strong excitonic binding energy (100-110 meV in the UV-visible range), and high optical transparency (> 80 \%), making it a promising host for blue-emitting nanostructures \cite{matz1993band, ferh1996elec, vett2019anal}. Bulk CuBr typically exhibits weak emission intensity at room temperature \cite{vija2017high}. Previous studies on $\gamma$-CuBr single crystals have shown that free-exciton emission is extremely temperature dependent, with the intensity at 77 K being $\sim$3500 times higher than at room temperature, underscoring significant thermal quenching that necessitates strategies to enhance room-temperature luminescence \cite{zhou2023grow}. CuBr offers enhanced optical properties; however, when confined to the QD regime, surface-related defects can limit emission efficiency. Despite these prospects, investigations into doped CuBr QDs remain unexplored. 

\hspace{5mm} In this work, we developed Zn:CuBr QDs through a cost-effective, room-temperature supersaturated recrystallization (SR) method with precise control over nucleation kinetics and delivering uniform size distributions, avoiding rapid supersaturation and polydispersity by a suitable choice of the solvent. Zn doping effectively passivates surface defects and bromine vacancies through stronger Zn-Br bonding, thereby suppressing nonradiative recombination and dramatically enhancing optical properties.
   
\section*{\normalsize Experimental details}

\subsubsection*{ Materials  }
\hspace{5mm} All the required chemicals such as Copper Bromide (CuBr, Thermo Fisher, 99.99 \% ), Zinc Bromide (ZnBr$_{2}$, Thermo Fisher, 99.99 \%), 1,4-dioxane (anhydrous 99.8 \%), Toluene  (anhydrous 99.8 \%), \textit{n}-Hexane (analytical reagent 98.9 \%), Oleyl Amine (OAm) and Oleyic Acid (OA) of analytical grade (purity > 99 \% ) were purchased from Sigma Aldrich and used without further purification.

\subsubsection*{\normalsize  Synthesis of CuBr QDs and Sn-doped CuBr QDs }
\hspace{5mm}
Zn:CuBr QDs were synthesized using the supersaturated recrystallization (SR) method. For this, 0.4 mmol of CuBr and 0.04 mmol of ZnBr$_2$ were dissolved in 10 mL of the polar aprotic solvent 1,4-dioxane and stirred at room temperature for 2 h to form a clear precursor solution. Oleylamine (25 $\mu$L) and oleic acid (50 $\mu$L) were added dropwise under continuous stirring to improve stability. Subsequently, 1 mL of the precursor solution was rapidly injected into 10 mL of toluene under vigorous stirring, yielding QDs. The QDs were collected by centrifugating at 7000 rpm for 5 min and re-dispersed in \textit{n}-hexane for storage.

\subsubsection*{\normalsize  Characterizations} 
\hspace{5mm} The XRD measurements were performed using a high-resolution X-ray diffractometer (Rigaku Smartlab) with Cu K$\alpha$ radiation (wavelength = 1.5406 $\angstrom$) scanned in the 2$\theta$ range 20$^{\circ}$- 70$^{\circ}$ (step size 0.02, scan rate 2$^{\circ}$/min). The HR-TEM microscope (Thermo Scientific, Talos F200X G2) was used to record high-resolution transmission electron microscopy (TEM) images. The elemental composition and chemical states of elements in pure and Zn-doped samples were analyzed by X-ray Photoelectron Spectroscopy (XPS) using a Scanning XPS Microprobe (PHI 5000 Versaprobe III). The ultraviolet-visible (UV-Vis) absorption spectra of QDs were recorded in the wavelength range of 200-800 nm with a spectrometer (Agilent Cary 5000). The photoluminescence emission spectra were obtained with a spectrofluorometer (Horiba PTI-400), and the lifetime of the sample was measured using a time-correlated single photon counting (TCSPC) lifetime fluorometer (Horiba Delta Flex).

\hspace{5mm}  
\section*{\normalsize   Results and Discussion}

\begin{figure}[!ht]
\centering
\includegraphics[width=1.0\textwidth]{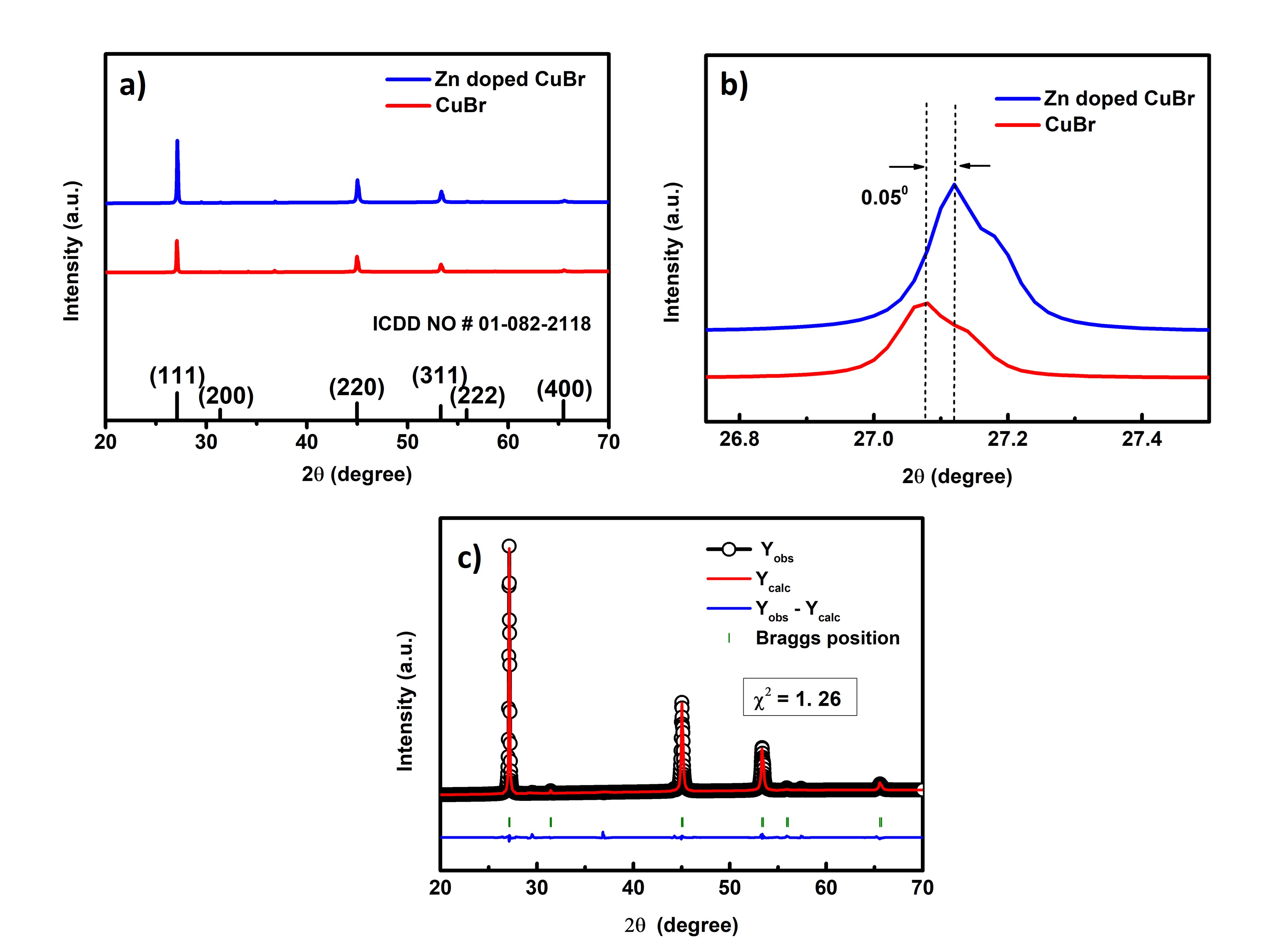} 
\caption{ a) The XRD pattern of CuBr and Zn:CuBr b) The shift of XRD pattern towards higher angle due to doping, c) Reitveld refinement profile of Zn:CuBr.} 
\label{fig:xrd}
\end{figure}

\hspace{5mm} Crystal structure and phase purity of the synthesized CuBr QDs and Zn-doped CuBr (Zn:CuBr) QDs are analyzed using X-ray diffraction (XRD) pattern shown in figure~\ref{fig:xrd}(a). The XRD pattern confirms that both the samples crystallize in a cubic structure with space group \textit{F4-3m}. The diffraction peaks for the pure CuBr QDs are observed at 2$\theta$ values 27.07$^{\circ}$, 31.39$^{\circ}$, 45.02$^{\circ}$, 53.31$^{\circ}$, 65.50$^{\circ}$. All peaks are well-matched with available standard diffraction information (ICDD PDF NO: 01-082-2118), and these peaks can be assigned to the planes  \textit{(200)}, \textit{(220)}, \textit{(311)}, \textit{(222)}, and \textit{(400)}. The d-spacing for the major peak is calculated as 3.29 \AA{} and the lattice parameter is determined to be 5.6980 \AA{}. For Zn:CuBr QDs, diffraction peaks appear at  27.12$^{\circ}$, 31.43$^{\circ}$, 44.97$^{\circ}$, 53.36$^{\circ}$, 65.56$^{\circ}$ corresponding to the planes \textit{(200)}, \textit{(220)}, \textit{(311)}, \textit{(222)}, and \textit{(400)}. The peaks shift to higher 2$\theta$ values after doping. The high-intense peak is shifted by 0.05$^{\circ}$ and is shown in figure~\ref{fig:xrd}(b). The d-spacing and lattice parameter for this peak are calculated as 3.284 \AA{} and 5.6896 \AA{}. The incorporation of Zn into CuBr leads to a decrease in d-spacing and lattice contraction, which can be attributed to the smaller ionic radius of $Zn^{2+}$ compared to $Cu^{+}$. This reduction in lattice parameter indicates that $Zn^{2+}$ ions are substituted for $Cu^{+}$ ions in the lattice rather than occupying interstitial positions. The Rietveld refinement of the Zn: CuBr sample is carried out using Fullprof software and shown in figure~\ref{fig:xrd}(c). 

The refinement was initiated by adopting structural parameters for pure CuBr, including phase, space group, lattice parameters, and atomic Wyckoff positions. The pseudo-Voigt function is applied to describe the peak shapes, and the quality of the refinement was verified through goodness-of-fit parameter $\chi^2$ and reliability factors ($R_{p}$, $R_{wp}$, and $R_{exp}$). In the refined structure, Cu and Zn atoms occupy the \textit{4a (0, 0, 0)} Wyckoff sites while Br atoms are located at \textit{4c (0.25, 0.25, 0.25)} positions. The obtained refinement parameters, $\chi^2$ = 1.26, $R_{p}$ = 7.64, $R_{wp}$ = 8.31, and $R_{exp}$ = 7.40, suggest an excellent match between the experimental and calculated diffraction profiles. Site occupancies for Cu, Br, and Zn were 0.03774, 0.04262, and 0.00443, respectively. These results confirm that the doped sample is also a single-phase crystalized in a cubic structure with space group \textit{F4-3m}. The lattice parameter obtained after refinement is 5.6914 \AA{}, which is consistent with the calculated value. The lattice contraction observed in the diffraction data is further confirmed by the refinement, and these results support the successful incorporation of Zn into the CuBr lattice.      

\begin{figure}[!ht]
\centering
\includegraphics[width=1.0\textwidth]{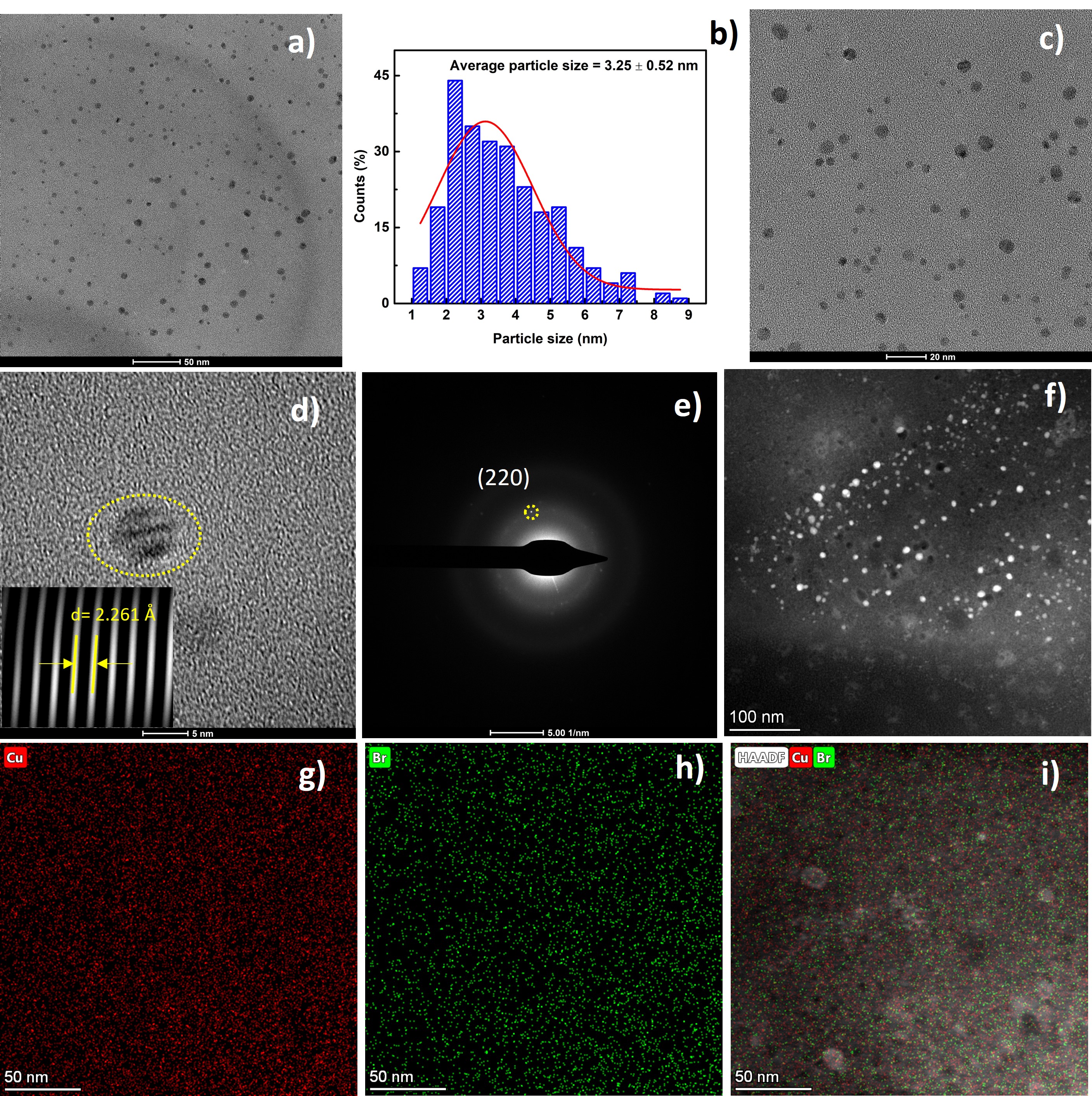} 
\caption{a) HR-TEM image in 50 nm, b) Particle size distribution histogram. HR-TEM image in c) 20 nm, d) 5 nm (Inverse FFT image in inset) e) SAED pattern, f) HAADF-STEM image and elemental mapping of g) Cu, h) Br, and i) CuBr of CuBr QDs. } 
\label{fig:temcu}
\end{figure}

\hspace{5mm} The high-resolution transmission electron microscopy (HR-TEM) is employed to investigate the morphology of the synthesized QDs. The HR-TEM images of CuBr QDs in 50 nm and 20 nm scale are shown in figure~\ref{fig:temcu}(a) and (d). The QDs are spherical in shape and uniformly dispersed. The particle size distribution histogram [(figure~\ref{fig:temcu}(b)] is obtained, and the average particle size is calculated by using  \textit{ImageJ} software as 3.25 $\pm$ 0.52 nm with a distribution of 1 nm to 9 nm.    
The high-resolution TEM image recorded at a 5 nm scale [Figure~\ref{fig:temcu}(d)] distinctly shows clear lattice fringes within a single QD. The inverse fast fourier transform (FFT) image [inset of Figure~\ref{fig:temcu}(d)] reveals a \textit{d}-spacing of 2.261 \AA{}, corresponding to the (\textit{220}) plane. The selected-area diffraction pattern (SAED) is shown in figure~\ref{fig:temcu}(e), further confirming the crystalline nature and showing diffraction spots corresponding to the \textit{(220)} plane with calculated \textit{d}-spacing of 2.248 \AA{}. The high-angle annular dark field scanning transmission electron microscopy (HAADF-STEM) image of CuBr QDs is given in figure~\ref{fig:temcu}(f) together with elemental mapping for Cu, Br, and CuBr, as shown in figure~\ref{fig:temcu}(g, h, i), confirming the uniform distribution of Cu and Br throughout the QDs.  
\begin{figure}[!ht]
\centering
\includegraphics[width=1.0\textwidth]{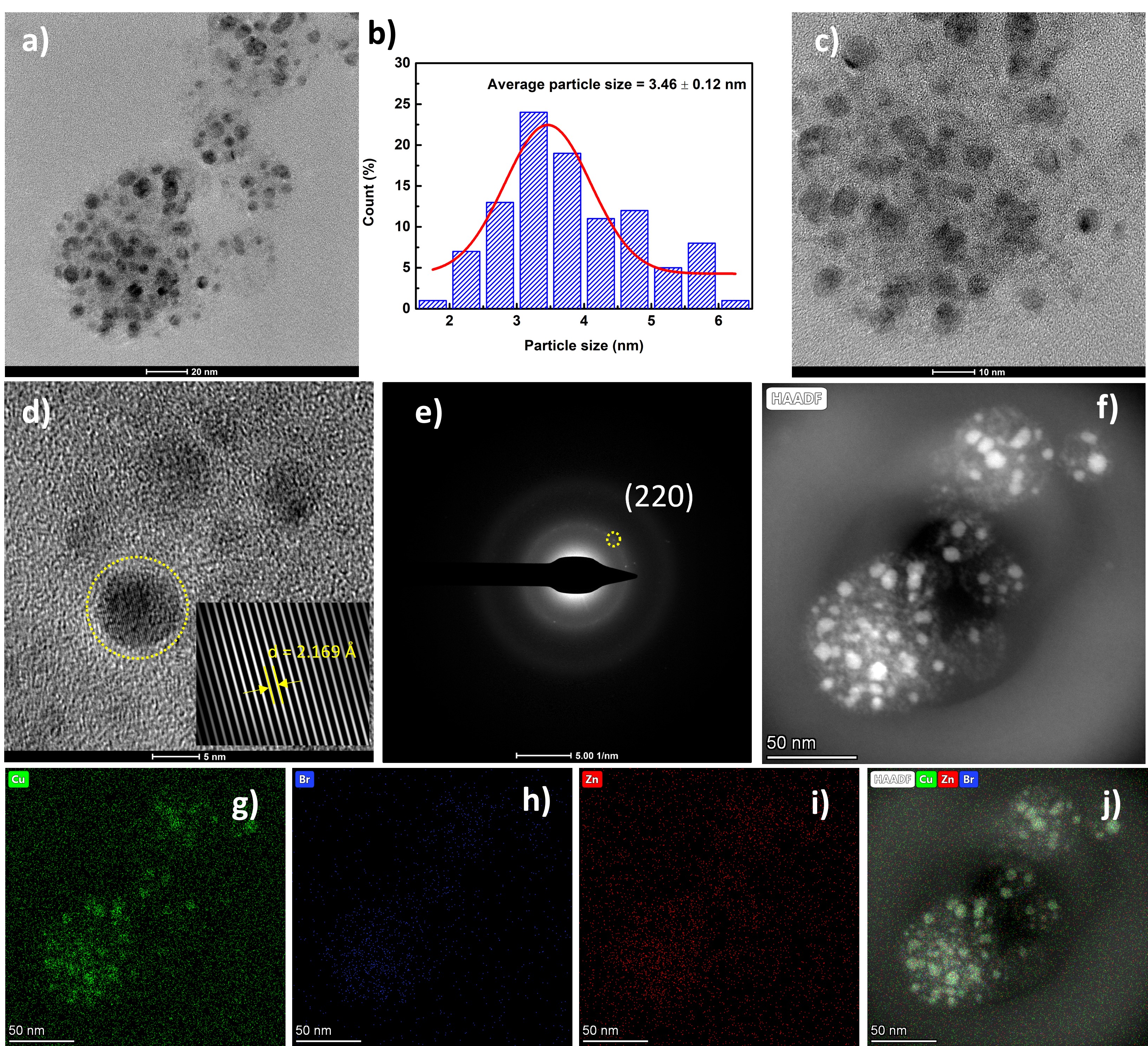} 
\caption{a) HR-TEM image in 20 nm, b) Particle size distribution histogram. HR-TEM image in c) 10 nm, d) 5 nm (Inverse FFT image in inset) e) SAED pattern, f) HAADF-STEM image and elemental mapping of g) Cu, h) Br, i) Zn, and j) Zn:CuBr of Zn:CuBr QDs. } 
\label{fig:temzn}
\end{figure}
HR-TEM images for Zn:CuBr QDs at 20 nm scale and 10 nm scale are shown in figure~\ref{fig:temzn}(a and c). The particle size is calculated as 3.46 $\pm$ 0.12 nm from the particle distribution histogram [(figure~\ref{fig:temzn}(b)]. The \textit{d}-spacing is calculated as 2.169 \AA{} for \textit{(220)} plane from the lattice fringes visible in the TEM image shown in figure~\ref{fig:temzn}(d) at 5 nm scale. This reduced \textit{d}-spacing by 0.092 \AA{} relative to undoped QDs again confirms the lattice contraction upon Zn doping. The diffraction spot obtained in the SAED pattern in figure~\ref{fig:temzn}(e) corresponds to \textit{(220)} reflection, and the \textit{d} spacing is calculated as 2.169\AA{}. The HAADF-STEM image and elemental mapping of the Cu, Br, Zn, and Zn:CuBr are shown in figure~\ref{fig:temzn}(f)-(j). These results confirm the successful incorporation of Zn into the CuBr lattice.

\hspace{5mm} 
\begin{figure}[!ht]
\centering
\includegraphics[width=1.0\textwidth]{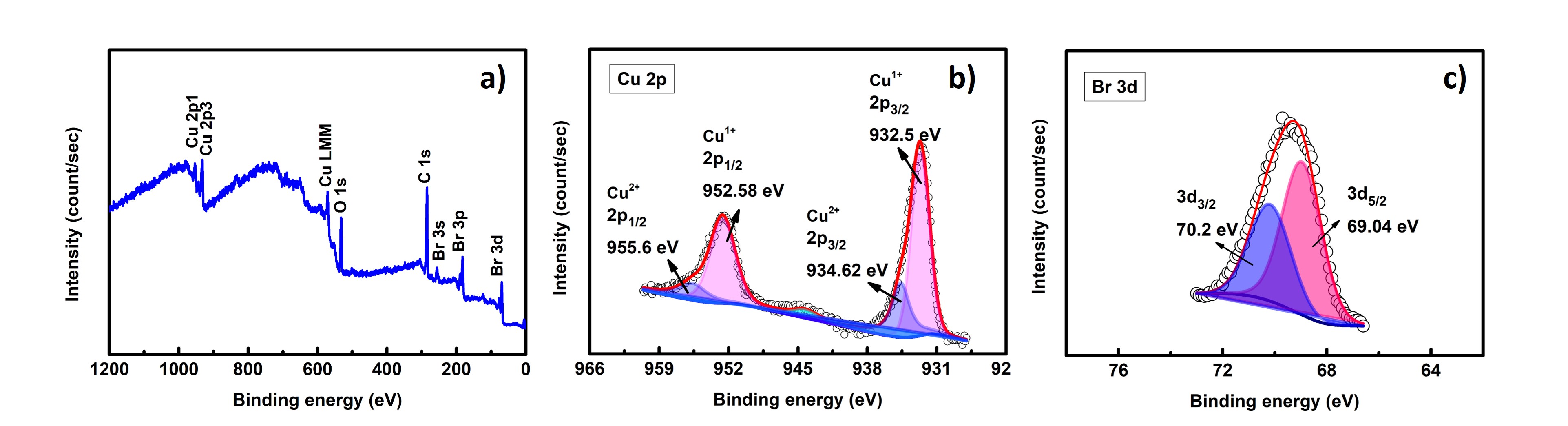} 
\caption{a) The high resolution survey spectra, Elemental scan of b) Cu 2p, c) Br 3d of CuBr QDs.} 
\label{fig:xps1}
\end{figure}
\hspace{5mm}Figure~\ref{fig:xps1} presents the X-ray photoelectron spectroscopy (XPS) analysis of pristine CuBr QDs, performed to investigate their elemental composition and chemical states. The wide-scan survey spectrum [(figure~\ref{fig:xps1}(a)] reveals prominent signals corresponding to Cu and Br, with no additional impurity-related peaks, confirming the formation of CuBr QDs with high chemical purity.
The high-resolution Cu 2p spectrum [(figure~\ref{fig:xps1}(b)] displays two distinct spin-orbit components located at 932.5 eV and 952.58 eV, corresponding to the Cu 2p$_{3/2}$ and Cu 2p$_{1/2}$ levels, respectively. The observed spin-orbit separation of approximately 20.08 eV is characteristic of copper in the +1 oxidation state, indicating that Cu$^{+}$ is the dominant species in the CuBr QDs. Additionally, weak features at higher binding energies of 934.62 eV and 955.6 eV, corresponding to the Cu 2p${3/2}$ and 2p${1/2}$ core levels, are indicative of Cu$^{2+}$, suggesting minor surface oxidation. The Br 3d core-level spectrum [(figure~\ref{fig:xps1}(c)] consists of two well-resolved peaks centered at 69.04 eV and 70.2 eV, which are assigned to the Br 3d$_{5/2}$ and Br 3d$_{3/2}$ states, respectively. The corresponding spin-orbit splitting of 1.16 eV is consistent with Br$^{-}$ ions chemically bonded within the CuBr lattice\cite{bies2017adva}.
\hspace{5mm} 
\begin{figure}[!ht]
\centering
\includegraphics[width=1.0\textwidth]{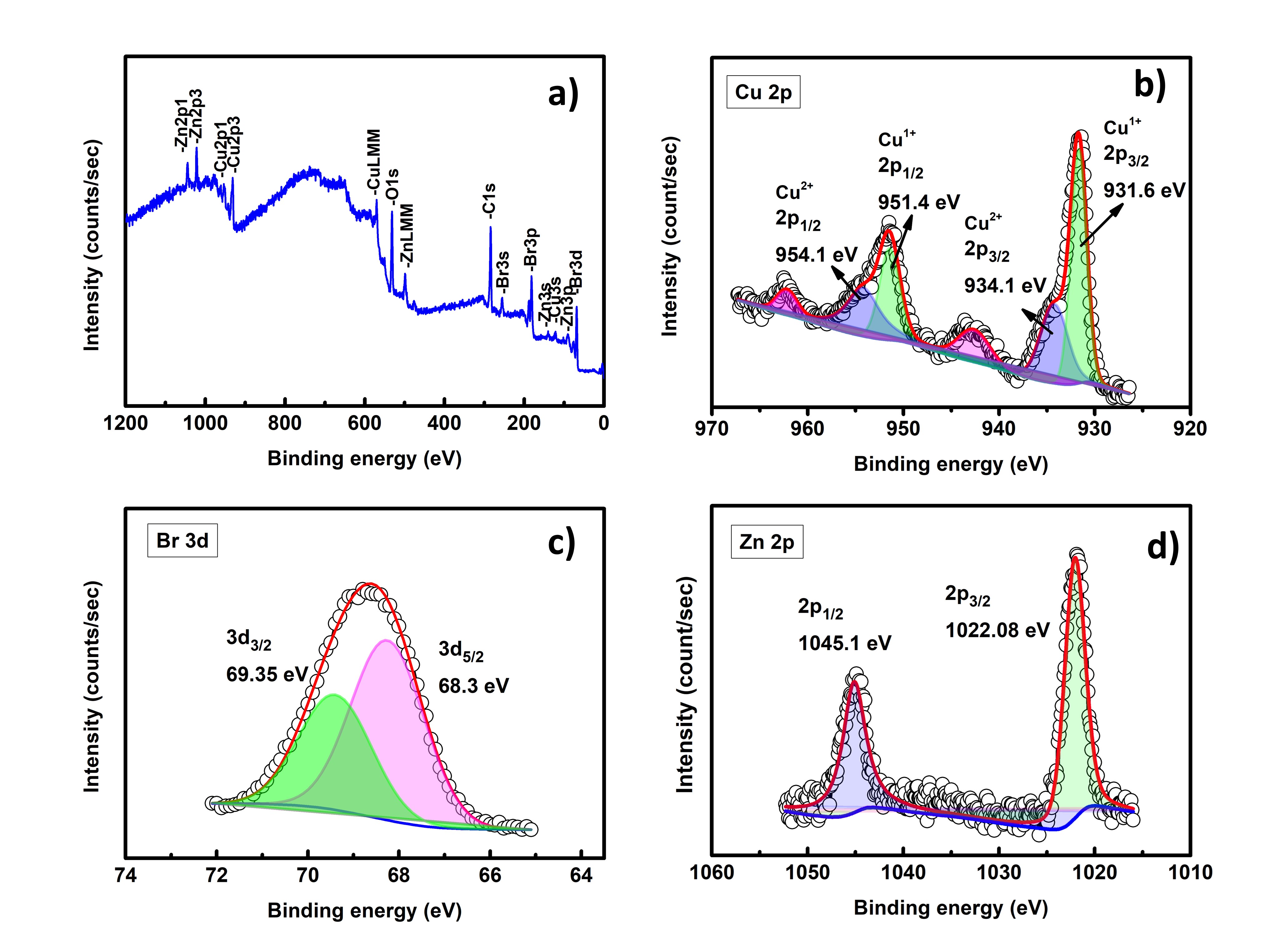} 
\caption{ a) XPS survey spectra, High resolution core spectra of b) Cu 2p, b) Br 3d, c) Zn 2p of Zn:CuBr QDs } 
\label{fig:xps2}
\end{figure}

Figure~\ref{fig:xps2} shows the XPS analysis of Zn-doped CuBr QDs. The survey spectrum [(figure~\ref{fig:xps2}](a) confirms the presence of Cu, Br, and Zn elements, with no detectable peaks from foreign elements, indicating successful Zn incorporation without contamination.
The high-resolution Cu 2p spectrum of the Zn-doped CuBr QDs (figure~\ref{fig:xps2}b) exhibits two dominant peaks at 931.6 eV (Cu 2p$_{3/2}$) and 951.4 eV (Cu 2p$_{1/2}$), with a spin-orbit separation of approximately 19.8 eV. These values are characteristic of Cu$^{+}$, demonstrating that the monovalent copper state is largely preserved after Zn doping. Additionally, weak peaks at 934.1 eV and 954.1 eV, corresponding to the Cu 2p${3/2}$ and 2p${1/2}$ core levels, are characteristic of the Cu$^{2+}$ oxidation state, indicating slight surface oxidation. Furthermore, distinct shake-up satellite peaks appear at 942.46 eV and 962.4 eV, which are typical fingerprints of Cu$^{2+}$ species \cite{yuan2020engi}.
The Br 3d spectrum of the Zn-doped sample (Figure~\ref{fig:xps2}c) shows two peaks at 68.3 eV and 69.35 eV, assigned to the Br 3d$_{5/2}$ and Br 3d$_{3/2}$ levels, respectively. The observed spin-orbit splitting of approximately 1.05 eV agrees well with reported values for Br$^{-}$ ions, confirming that the Cu-Br bonding environment remains intact upon Zn doping. Figure~\ref{fig:xps2}(d) presents the high-resolution Zn 2p spectrum of the Zn-doped CuBr QDs. Two distinct peaks are observed at 1022.08 eV and 1045.1 eV, corresponding to the Zn 2p$_{3/2}$ and Zn 2p$_{1/2}$ states, respectively. The measured spin-orbit separation of 23.0 eV is characteristic of Zn$^{2+}$, confirming that Zn is present in the divalent oxidation state. The absence of additional Zn-related features suggests that Zn is incorporated into the CuBr lattice without the formation of secondary zinc-based phases.
\hspace{5mm}    
\begin{figure}[!ht]
\centering
\includegraphics[width=1.0\textwidth]{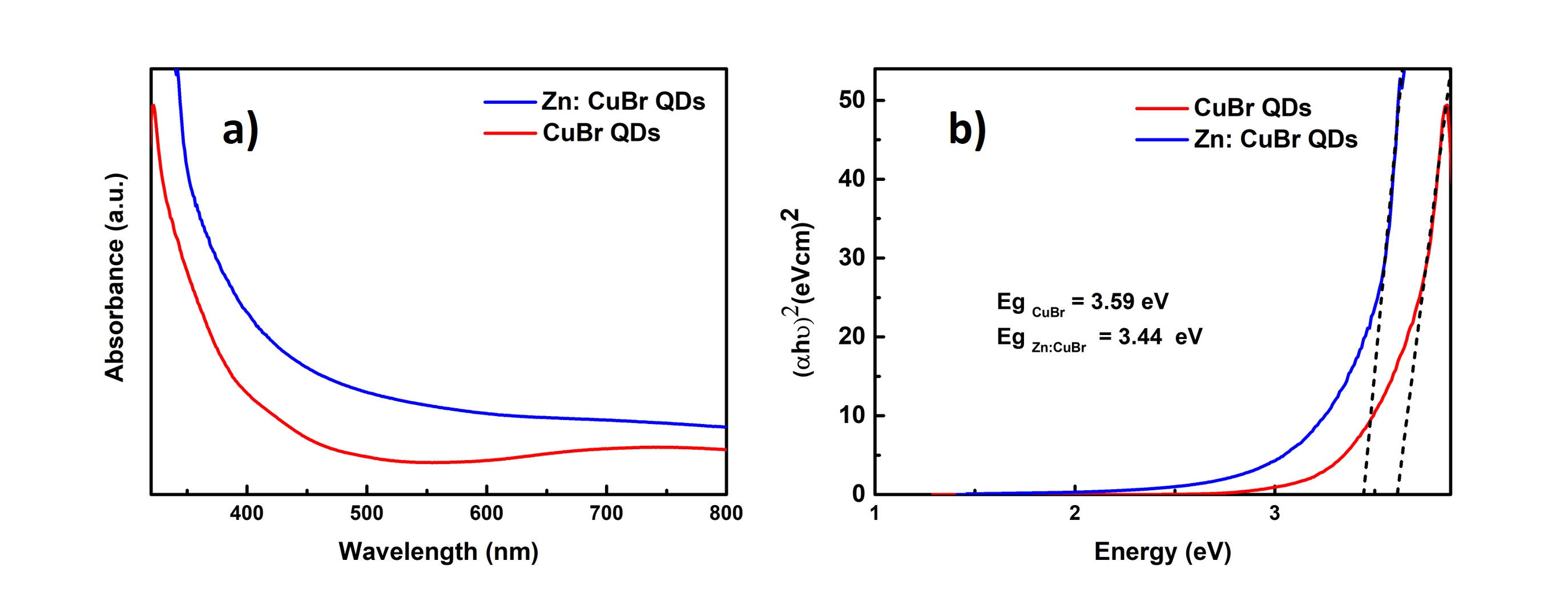} 
\caption{a) The absorbance spectra and b) Tauc plot for bandgap energy calculation of CuBr QDs and Zn:CuBr QDs.} 
\label{fig:uv}
\end{figure}

The optical properties of the synthesized QDs were investigated using UV-Vis absorption spectroscopy (UV-Vis). The absorbance spectra of CuBr QDs and Zn-doped CuBr QDs are shown in figure~\ref{fig:uv}(a). For CuBr QDs, the absorbance edge is near 320 nm, which is significantly blue-shifted compared to the bulk CuBr absorption edge at around 380 nm \cite{xiab2024aself}, indicating pronounced size-dependent quantum confinement effects. In addition, a broad absorption feature in the 387-400 nm range is attributed to band-edge excitonic absorption rather than a distinct, sharp peak. In bulk CuBr, two closely spaced excitonic transitions are well known at approximately 3.1 eV ($Z_{1,2}$) and 3.17 eV ($Z_{3}$) \cite{stef2026surf}. However, in the QD regime, these transitions merge into a single broadened excitonic band due to the combined effects of quantum confinement, particle size distribution, and surface-related interactions. Furthermore, the extended absorption tail toward longer wavelengths is indicative of band-edge disorder and defect-induced Urbach states. The absorbance spectra edge of CuBr after Zn doping is red-shifted to 341 nm, attributable to reduced quantum confinement arising from an enlarged particle size\cite{canv2020infl}.

The optical bandgap for the synthesised samples were calculated using Tauc relation; $\alpha h \nu = A(h\nu-E_g)^n $  where $E_{g}$ represents optical bandgap, `\textit{h}' is Plank's constant, $\alpha$ is absorption coefficient, `\textit{A}' is the band tailoring parameter and the index `\textit{n}' reflects the optical transition type (1/2 for direct allowed, 2 for indirect allowed, 3 for direct forbidden, or 3/2 for indirect forbidden transitions). Since CuBr has a direct bandgap, n=1/2 was used. As shown in figure~\ref{fig:uv}(b), $E_{g}$ was obtained by extending the linear region of the $(\alpha\, h\nu)^2$ vs. $h\nu$ plot to intersect the \textit{x}-axis.
The calculated bandgap energy for CuBr QDs is 3.59 eV, significantly higher than the 3.1 eV for bulk CuBr \cite{mfer1996elec}. With an exciton-Bohr radius of 1.25 nm, the synthesized QDs exceed this size, placing them in the weak confinement regime where bandgap enhancement arises from quantum confinement effects. Zn incorporation is accompanied by a decrease in the bandgap from 3.59 eV in pristine QDs to 3.44 eV in the doped samples. This bandgap narrowing may be influenced by several factors, including defect formation, lattice distortion, dopant-related electronic states, and reduced quantum confinement due to particle-size variations. Among these possible mechanisms, changes in the effective size of CuBr QDs induced by $Zn^{2+}$ doping may play a significant role in the observed bandgap reduction\cite{vale1998dyna, huon2024eudo}.
The size-induced modification of the bandgap of CuBr QD was quantitatively evaluated using the effective mass approximation, in which the bandgap of a semiconductor QD is expressed as the sum of the bulk bandgap, a quantum confinement term, and a Coulomb interaction term according to the Brus model:
\begin{equation}
E_g^{QD} = E_g^{\mathrm{bulk}}+ \frac{\hslash^2 \pi^2}{2R^2}
\left( \frac{1}{m_e^*} + \frac{1}{m_h^*} \right)
- \frac{1.8 e^2}{4\pi \varepsilon_0 \varepsilon R}
\end{equation}
where $E_g^{QD}$ and $E_g^{\mathrm{bulk}}$ are the bandgap energies of QD and Bulk CuBr, respectively. For CuBr, ($E_{g}^{bulk}$= 3.1 eV, effective mass of electron, $m_{e}^*$= 0.22, effective mass of hole $m_{h}^*$= 1.11, dielectric constant $\varepsilon$= 5.2, $\varepsilon(0)$ = 7.9, \textit{e} is the elementary charge and the experimentally obtained average QD radius, R = 1.625 nm. Brus equation calculations predict the QD bandgap of 3.57 eV. This value closely matches the Tauc plot results from UV-Vis spectra, which will be discussed in the following section, and quantitatively accounts for the 0.49 eV increase from the bulk, a signature of weak quantum confinement effects.

\begin{figure}[!ht]
\centering
\includegraphics[width=1.0\textwidth]{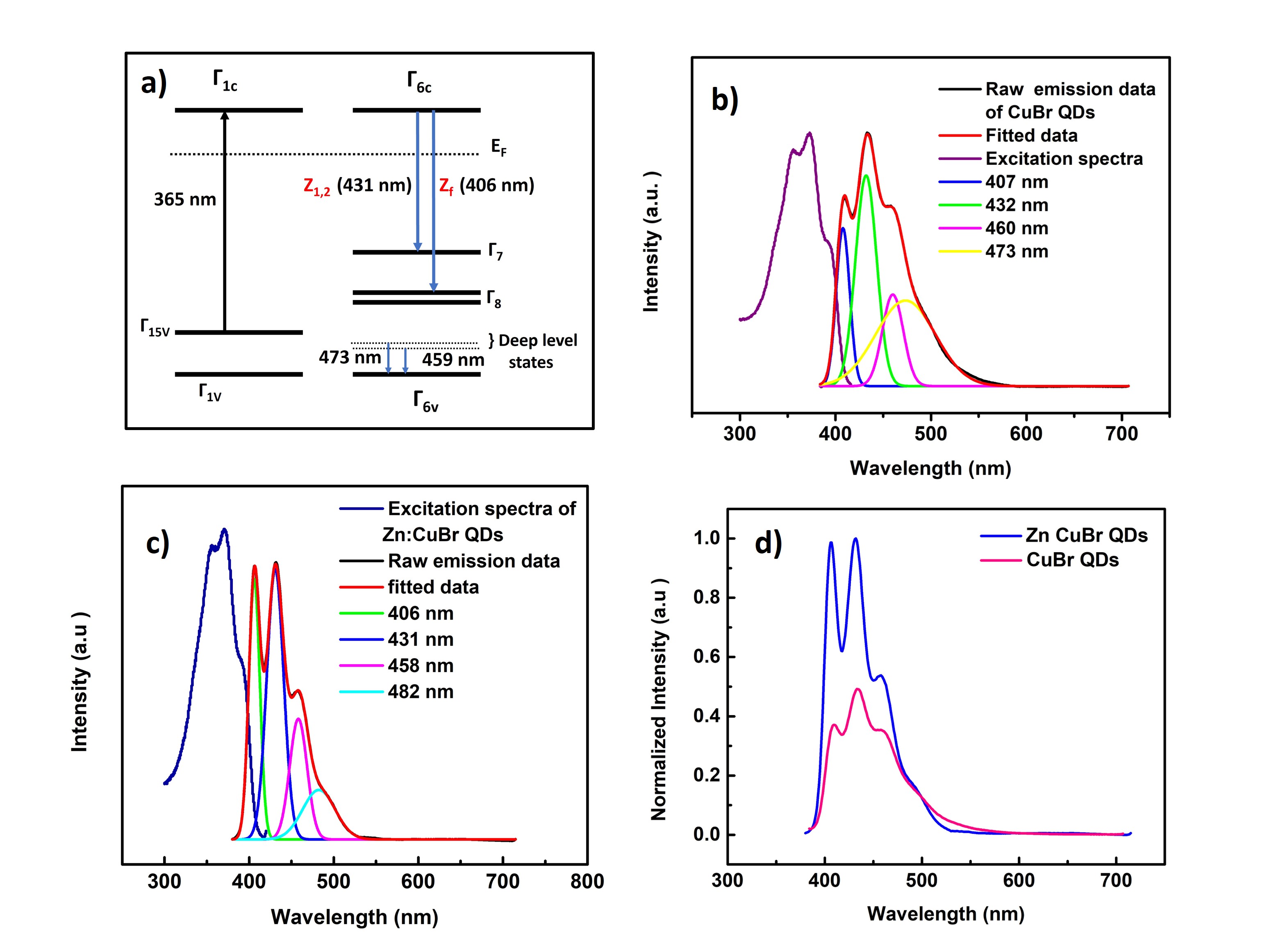} 
\caption{a) Schematic energy level diagram for CuBr QDs. Excitation spectra and deconvoluted photoluminescence emission spectra of b) CuBr QDs. c) Zn:CuBr QDS. d) Normalized photoluminescence spectra of pure and Zn-doped QDs showing enhanced emission intensity after Zn doping.} 
\label{fig:pl}
\end{figure}

The electronic structure of CuBr QD is governed by strong hybridization between Cu 3d states and halogen \textit{p} orbitals, which fundamentally distinguishes copper halides from conventional III-V and II-VI semiconductors, where the 3d levels remain largely localized. In CuBr, the Cu 3d states lie energetically close to the halogen \textit{p} states, leading to substantial orbital mixing that strongly influences the excitonic landscape. At the $\varGamma$, the excitonic states follow the ascending energy order $Z_{f}$, $Z_{1,2}$, and $Z_{3}$. Cu 3d orbitals split into \textit{e} and $t_{2}$ components within the Zinc-Blende CuBr lattice with $T_{d}$ symmetry. The $t_{2}$ states strongly hybridize with halogen \textit{p} orbitals to form an antibonding $\varGamma_{15v}$ state at the valence band maximum, along with a corresponding bonding state at lower energy. The \textit{e}-type Cu 3d orbitals give rise to a nonbonding $\varGamma_{12v}$ state located well below the valence band edge. The conduction band minimum originates from antibonding interactions between Cu 4s and halogen \textit{s} orbitals and is represented by the $\varGamma_{1c}$ 
state. Spin-orbit coupling further lifts the degeneracy of the $\varGamma_{15v}$ state, resulting in $\varGamma_{8}$ and $\varGamma_{7}$ subbands. Optical transitions in CuBr QDs are therefore dominated by recombination between the lowest conduction band state ($\varGamma_{6}$ and holes in $\varGamma_{8}$ and $\varGamma_{7}$ valence subbands, corresponding to the $Z_{1,2}$ and $Z_{3}$ excitonic transitions  \cite{gold1977band, mfer1996elec, serr2002elec,haoy2022band} illustrated in the energy-level diagram figure~\ref{fig:pl}(a). 

\hspace{5mm} Figure~\ref{fig:pl}(b) shows the Gaussian-deconvoluted photoluminescence (PL) spectrum of CuBr QDs measured under 365 nm excitation. Four emission components contributing to the blue emission are resolved at approximately 407, 432, 460, and 473 nm, with corresponding full-width half-maximum (FWHM) values of 16.56, 26.10, 25.89, and 73.54 nm, respectively. The high-energy emissions at 407 and 432 nm are assigned to near-band-edge recombination, originating from the free exciton ($Z_{f}$) and bound exciton ($Z_{1,2}$) states, respectively. Conversely, the lower-energy bands centered at 460 and 473 nm correspond to deep-level radiative transitions associated with defects: specifically, bromine deficiencies and copper vacancies, respectively. These defect-related emissions are frequently suppressed because they correspond to slow, non-radiative-assisted recombination pathways that can reduce luminescence efficiency \cite{zhou2023grow}. Quantitative analysis of the integrated peak areas reveals that band-edge emission accounts for $\sim$48 \% of the total PL intensity, while defect-assisted channels contribute $\sim$52 \%. The dominance and broad linewidth of the defect-related emission indicate the presence of non-ideal surface or lattice states that act as competing recombination pathways, thereby degrading emission purity. This interpretation is further supported by the excitation spectrum, which exhibits features at 355 nm corresponding to hot-carrier excitation, a dominant absorption at 373 nm associated with band-edge transitions, and a lower-energy feature at 394 nm attributed to shallow surface states that facilitate carrier trapping. Consequently, suppression of these defect states through suitable passivation or compositional control is necessary to enhance the relative contribution of band-edge recombination for optoelectronic applications. 
\hspace{5mm} 

The excitation spectrum of Zn-doped CuBr QDs (figure~\ref{fig:pl}(c)) shows peaks at 355, 373, and 392 nm, closely mirroring the pristine sample and confirming preservation of core electronic structure and band-edge absorption characteristics. The PL response is markedly modified upon $Zn^{2+}$ incorporation, as shown in Figure~\ref{fig:pl}(c). Zinc effectively passivates surface defects, owing to stronger Zn-Br bonding than Cu-Br, thereby reducing the density of trap states and suppressing non-radiative recombination pathways. As a result, excitonic recombination becomes the dominant emission mechanism. The PL spectrum of Zn-doped CuBr QDs exhibits emission peaks at 406, 431, 458, and 482 nm, with FWHM values of 15.41, 23.44, 24.24, and 45.30 nm, respectively. 

Importantly, quantitative area analysis indicates that band-edge emission increased to $\sim$67 \% of the total PL intensity, accompanied by a corresponding reduction of defect-assisted contributions to $\sim$33 \%, reflecting a substantial mitigation of trap-mediated recombination rather than its complete elimination. The pronounced narrowing of the defect-associated emission band from 73.54 to 45.30 nm, together with its slight red-shift from 473 to 482 nm, provides strong evidence for effective defect passivation, likely arising from surface enrichment and partial lattice incorporation due to Zn doping.   
Figure~\ref{fig:pl}(d) presents the normalized PL spectra of pristine and Zn-doped CuBr QDs, under identical excitation conditions, clearly illustrating the suppression of low-energy defect-related emission and the redistribution of spectral weight toward near-band-edge excitonic recombination upon Zn incorporation. Collectively, these results demonstrate that Zn doping efficiently mitigates defect-mediated recombination pathways, thereby improving the emission purity and optical quality of CuBr QDs, a highly desirable feature for blue-emitting optoelectronic applications.

Overall, Zn-doped CuBr QDs provide a simpler, lead-free alternative that delivers better blue emission than the pure and improved color purity through intrinsic defect passivation, highlighting their potential as efficient, environmentally benign emitters for optoelectronic applications.

\begin{figure}[!ht]
\centering
\includegraphics[width=1.0\textwidth]{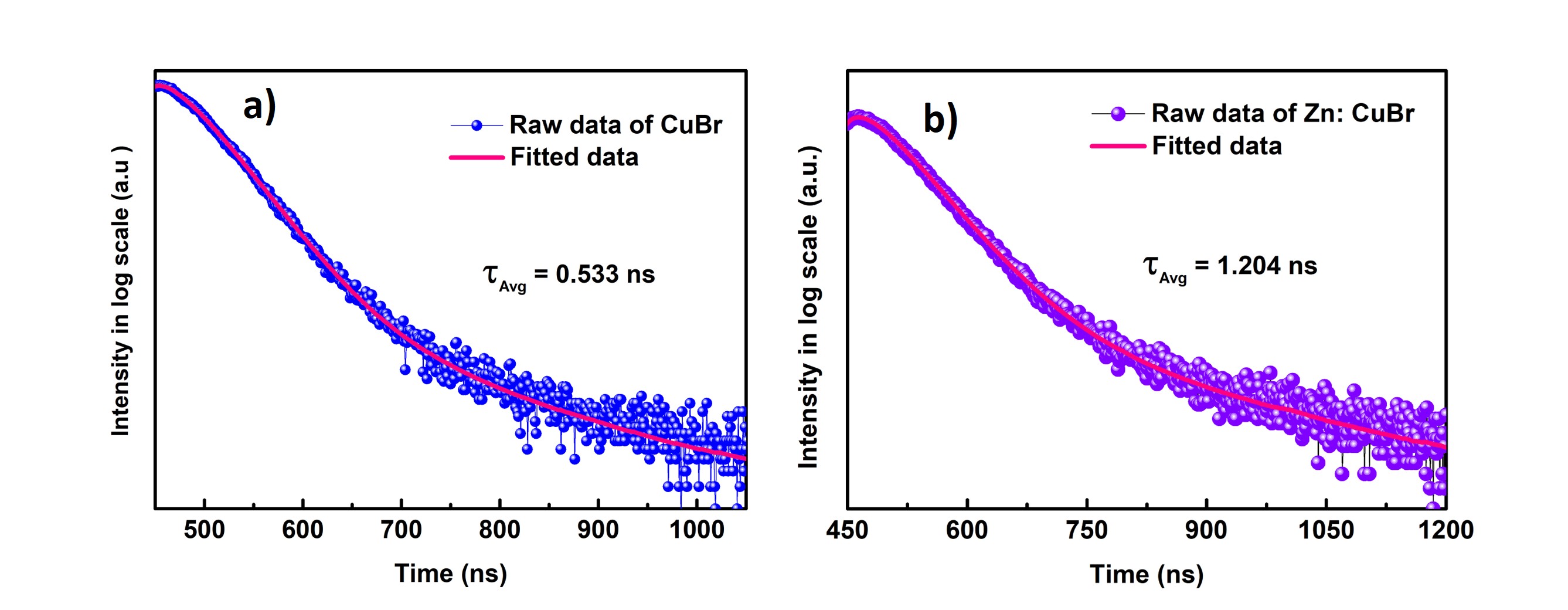} 
\caption{ TCSPC decay profile of a) CuBr QDs, b) Zn:CuBr QDs.} 
\label{fig:decay}
\end{figure}
The PL decay dynamics of CuBr and Zn-doped CuBr QDs were examined using time-correlated single-photon counting (TCSPC) measurements under 365 nm excitation, monitoring emission near 431 nm to probe the recombination mechanisms. The decay profile [figure~\ref{fig:decay}(a)] follows a tri-exponential function for pure CuBr QDs, and is fitted with the equation,
\begin{equation}
 A(t)= A_0+ A_1 e^{(-t/\tau_1)} + A_2 e^{(-t/\tau_2)} + A_3 e^{(-t/\tau_3)}
\end{equation}

where ${A_0}$ is a constant, ${A_1}$, ${A_2}$, ${A_3}$ are weighted amplitudes and ${\tau_1}$, ${\tau_2}$, ${\tau_2}$ corresponds to the fitted lifetime component of decay. The fitting parameters are obtained as, ${A_1}$ = 14.39 \%, ${A_2}$ = 79.70 \% and ${A_3}$ = 5.91 \%. The corresponding lifetimes are  ${\tau_1}$ = 0.15 ns, ${\tau_2}$ = 0.9 ns, ${\tau_3}$ = 3.51 ns. Here, the fastest decay component arises from surface-related non-radiative recombination, the intermediate lifetime links to band-edge radiative processes, and the dominant slow component arises from trap-state mediated charge carrier recombination\cite{jana2022inve}.
The intensity-weighted average lifetime of CuBr QDs is calculated by using the equation,
\begin{equation}
\tau _{avg}=\frac{A_1\ \tau_1^2 +  A_2\ \tau_2^2 + A_3\ \tau_3^2}{A_1\ \tau_1 + A_2\ \tau_2 + A_3\ \tau_3}
\end{equation}
as 0.533 ns. The fitting parameters for Zn:CuBr QDs are obtained as ${A_1}$ = 70.32 \%, ${A_2}$ = 24.51 \%, and ${A_3}$ = 5.17 \%. The corresponding lifetimes are  ${\tau_1}$ = 1.03 ns, ${\tau_2}$ = 1.71 ns, ${\tau_3}$ = 6.06 ns. The average lifetime is calculated as 1.204 ns. Zn doping shifts the decay toward longer lifetimes, with a reduced fast component and enhanced intermediate or slow contributions, indicating suppressed non-radiative pathways and improved radiative efficiency. 

\begin{figure}[!ht]
\centering
 \includegraphics[width=1.0\textwidth]{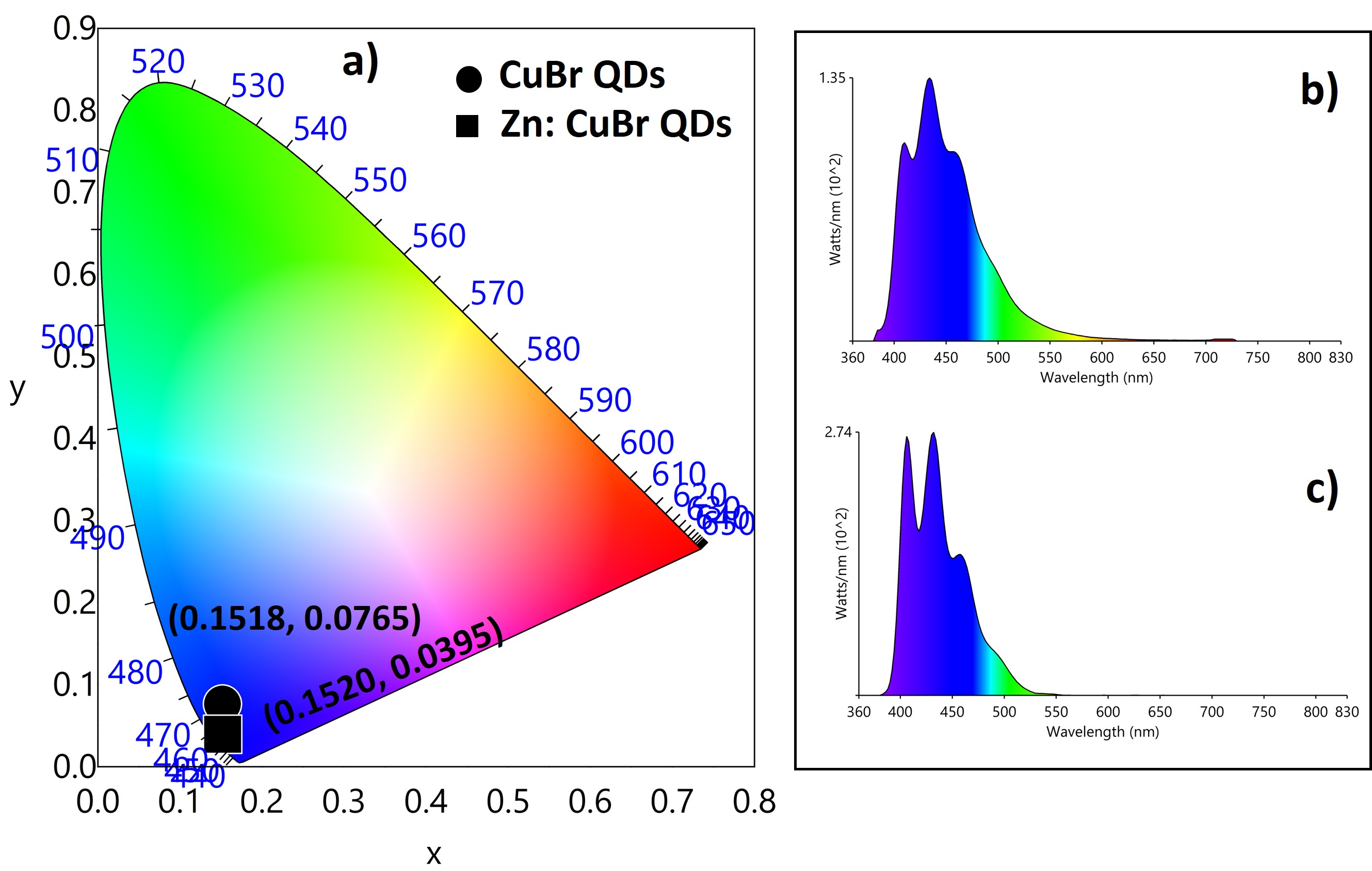} 
\caption{a) CIE diagram, spectrum plot of b) CuBr QDs, c) Zn:CuBr QDs} 
\label{fig:cie}
\end{figure}

The colorimetric properties of pure and Zn-doped CuBr QDs were evaluated using the CIE 1931 chromaticity diagram [figure~\ref{fig:cie}(a)], constructed from their respective PL spectra with the corresponding spectrum plot in [figure~\ref{fig:cie}(b) and (c)]. 
The color purity is calculated using the equation,
\begin{equation}
  CP = \frac{\sqrt{(x-x_i)^2+(y-y_i)^2}}{\sqrt{(x_d-x_i)^2+(y_d-y_i)^2}}\times100\%
\end{equation}
where ($x_{i}$,$y_{i}$) represent the white point illuminant coordinates, and ($x_{d}$,$y_{d}$) denotes the dominant wavelength boundary on the spectral locus.
The pure CuBr QDs exhibit CIE  coordinates (0.1518, 0.0765) with 89.6 \% color purity, while Zn-doping shifts these to (0.1520, 0.0395), achieving  96.4 \% purity. This improvement in monochromaticity, reflected by a deeper shift into the blue region, is attributed to the suppression of defect-related broadband emission upon doping, resulting in emission closer to the spectral locus boundary and improved color quality for display and LED applications.

\section*{\normalsize  Conclusion}
\hspace{5mm} 
We successfully synthesized blue-emitting CuBr QDs and passivated them with Zn via a simple, cost-effective supersaturated recrystallization method at room temperature. Structural characterization via XRD confirmed the purity and crystallinity of the cubic phase, while Rietveld refinement and XPS analysis confirmed the successful incorporation of Zn$^{2+}$ into the CuBr lattice. HR-TEM analysis reveals uniform particle sizes of 3.25 $ \pm$ 0.52 nm for pure CuBr QDs and 3.46$ \pm$ 0.12\ nm for Zn-doped samples. Optical characterization shows that CuBr QDs exhibit blue emission, dominated by defect-assisted recombination arising from bromine vacancies and surface trap states. Upon Zn doping, the bandgap decreases from 3.59 eV to 3.44 eV, due to weakened quantum confinement resulting from a slight increase in particle size and modifications to defect-related electronic states. More importantly, Zn$^{2+}$ ions effectively passivate bromine vacancies via strong Zn-Br interactions, thereby suppressing defect-related broadband emission and narrowing the emission spectrum. This increases the contribution of band-edge excitonic recombination from 48\% to 67\%. TCSPC analysis further supports the defect passivation mechanism, revealing a substantial increase in the average carrier lifetime from 0.53 ns to 1.20 ns due to reduced non-radiative recombination pathways. In addition, Zn:CuBr QDs exhibit highly pure deep-blue emission with a color purity of 96.4 \% and stable CIE chromaticity coordinates of (0.1520, 0.0395). Overall, these results establish Zn doping as a simple, scalable, and environmentally benign approach to mitigate intrinsic defects in CuBr QDs.

\section*{Acknowledgment}
Aryamol Stephen thanks the Department of Science and Technology (DST), Government of India, for providing a Women in Science and Engineering (WISE) fellowship [DST/WISE- PhD/PM/2024/30]. The authors acknowledge the Indian Institute of Technology (Indian School of Mines), Dhanbad, Jharkhand, India, for providing various analytical and characterization facilities.

\section*{CRediT authorship contribution statement}

\textbf{Aryamol Stephen:} Conceptualization,  Data curation, Formal analysis, Methodology, Writing--original draft. \textbf{Ruchi Kumari:} Data curation, Formal analysis. \textbf{Saparja Roy:} Data curation, Formal analysis. \textbf{P. Jayaram:} Formal analysis, and Writing--Review \& editing,
\textbf{A. Biju:} Conceptualization, Formal analysis, Investigation,  Methodology, Supervision, Validation, and Writing--Review \& editing.
\textbf{P. M. Sarun:} Conceptualization, Formal analysis, Investigation,  Methodology,  Resources, Supervision, Validation, and Writing--Review \& editing.

\section*{Declaration of competing interest}

The authors declare that they have no known competing financial interests or personal relationships that could have appeared to influence the work reported in this paper.

\section*{Data availability}

Data will be made available on reasonable request to the corresponding author.

\newpage
\singlespacing

\begin{thebibliography}{10}
\expandafter\ifx\csname url\endcsname\relax
  \def\url#1{\texttt{#1}}\fi
\expandafter\ifx\csname urlprefix\endcsname\relax\def\urlprefix{URL }\fi
\expandafter\ifx\csname href\endcsname\relax
  \def\href#1#2{#2} \def\path#1{#1}\fi

\bibitem{renj2025deep}
J.~Ren, J.~Liu, B.~Wei, W.~Zhang, L.~Edman, J.~Wang, Deep-blue and
  narrowband-emitting carbon dots from a sustainable precursor for random
  lasing, ACS Applied Nano Materials 8~(5) (2025) 2472--2480.

\bibitem{sree2025rece}
S.~Sreejith, J.~Ajayan, N.~U. Reddy, M.~Manikandan, S.~Umamaheswaran, N.~R.
  Reddy, Recent advancements in high efficiency deep blue organic light
  emitting diodes, Micro and Nanostructures 200 (2025) 208101.

\bibitem{tang20234dop}
H.~Tang, Z.~Liu, H.~Zhang, X.~Zhu, Q.~Peng, W.~Wang, T.~Ji, P.~Zhang, Y.~Le,
  A.~N. Yakovlev, et~al., 4{D} optical information storage from
  {L}i{G}a$_{5}${O}$_{8}$: {C}r$^{3+}$ nanocrystal in glass, Advanced Optical
  Materials 11~(15) (2023) 2300445.

\bibitem{huat2024deep}
T.~Hua, X.~Cao, J.~Miao, X.~Yin, Z.~Chen, Z.~Huang, C.~Yang, Deep-blue organic
  light-emitting diodes for ultrahigh-definition displays, Nature Photonics
  18~(11) (2024) 1161--1169.

\bibitem{heho2025envi}
H.~He, S.~Deng, Y.~Liu, Environmentally friendly synthesis of quantum dots and
  their applications in diverse fields from the perspective of environmental
  compliance: A review, Discover Nano 20~(1) (2025) 132.

\bibitem{leew2020synt}
W.~Lee, C.~Lee, B.~Kim, Y.~Choi, H.~Chae, Synthesis of blue-emissive
  {InP/GaP/ZnS} quantum dots via controlling the reaction kinetics of shell
  growth and length of capping ligands, Nanomaterials 10~(11) (2020) 2171.

\bibitem{lixi2020cdse}
X.~Li, H.~Zhang, F.~Sun, {CdSe/ZnS} quantum dots exhibited nephrotoxicity
  through mediating oxidative damage and inflammatory response, Aging (Albany
  NY) 13~(8) (2020) 12194.

\bibitem{yili2000thep}
L.~Yi, Y.~Hou, H.~Zhao, D.~He, Z.~Xu, Y.~Wang, X.~Xu, The photo-and
  electro-luminescence properties of {Z}n{O}: {Z}n thin film, Displays 21~(4)
  (2000) 147--149.

\bibitem{cowl2010elec}
A.~Cowley, F.~O. Lucas, E.~Gudimenko, M.~Alam, D.~Danieluk, A.~Bradley,
  P.~McNally, Electroluminescence of $\gamma$-{CuBr} thin films via vacuum
  evaporation depositon, Journal of Physics D: Applied Physics 43~(16) (2010)
  165101.

\bibitem{liru2025mult}
R.~Li, J.~Zhao, Y.~Qiao, X.~Liu, S.~Mei, Multifunctional colloidal quantum
  dots-based light-emitting devices for on-chip integration, Nanomaterials
  15~(18) (2025) 1422.

\bibitem{zhan2023coll}
J.~Zhang, S.~Zhang, Y.~Zhang, O.~A. Al-Hartomy, S.~Wageh, A.~G. Al-Sehemi,
  Y.~Hao, L.~Gao, H.~Wang, H.~Zhang, Colloidal quantum dots: synthesis,
  composition, structure, and emerging optoelectronic applications, Laser \&
  Photonics Reviews 17~(3) (2023) 2200551.

\bibitem{piet2016spec}
J.~M. Pietryga, Y.-S. Park, J.~Lim, A.~F. Fidler, W.~K. Bae, S.~Brovelli, V.~I.
  Klimov, Spectroscopic and device aspects of nanocrystal quantum dots,
  Chemical Reviews 116~(18) (2016) 10513--10622.

\bibitem{wang2015brig}
A.~Wang, H.~Shen, S.~Zang, Q.~Lin, H.~Wang, L.~Qian, J.~Niu, L.~Song~Li,
  Bright, efficient, and color-stable violet {Z}n{S}e-based quantum dot
  light-emitting diodes, Nanoscale 7 (2015) 2951--2959.

\bibitem{zhou2015towa}
J.~Zhou, Y.~Yang, C.~Y. Zhang, Toward biocompatible semiconductor quantum dots:
  from biosynthesis and bioconjugation to biomedical application, Chemical
  Reviews 115~(21) (2015) 11669--11717.

\bibitem{kaga2020coll}
C.~R. Kagan, L.~C. Bassett, C.~B. Murray, S.~M. Thompson, Colloidal quantum
  dots as platforms for quantum information science, Chemical Reviews 121~(5)
  (2020) 3186--3233.

\bibitem{kirk2018find}
N.~Kirkwood, J.~O. Monchen, R.~W. Crisp, G.~Grimaldi, H.~A. Bergstein,
  I.~Du~Foss{\'e}, W.~Van Der~Stam, I.~Infante, A.~J. Houtepen, Finding and
  fixing traps in {II-VI and III-V} colloidal quantum dots: the importance of
  {Z}-type ligand passivation, Journal of the American Chemical Society
  140~(46) (2018) 15712--15723.

\bibitem{kimh2019emis}
H.-J. Kim, J.-H. Jo, S.-Y. Yoon, D.-Y. Jo, H.-S. Kim, B.~Park, H.~Yang,
  Emission enhancement of {C}u-doped {InP} quantum dots through double shelling
  scheme, Materials 12~(14) (2019) 2267.

\bibitem{makk2018fron}
M.~Makkar, R.~Viswanatha, Frontier challenges in doping quantum dots: synthesis
  and characterization, RSC Advances 8~(39) (2018) 22103--22112.

\bibitem{matz1993band}
R.~Matzdorf, J.~Skonieczny, J.~Westhof, H.~Engelhard, A.~Goldmann, Band
  structure and optical properties of {CuBr}: new photoemission results,
  Journal of Physics: Condensed Matter 5~(23) (1993) 3827.

\bibitem{ferh1996elec}
M.~Ferhat, A.~Zaoui, M.~Certier, J.~Dufour, B.~Khelifa, Electronic structure of
  the copper halides {CuCl}, {CuBr} and {CuI}, Materials Science and
  Engineering: B 39~(2) (1996) 95--100.

\bibitem{vett2019anal}
R.~Vettumperumal, J.~R. Jemima, S.~Kalyanaraman, R.~Thangavel, Analysis of
  electronic and optical properties of copper iodide ($\gamma$-{CuI}) by
  {TB-mBJ} method-a promising optoelectronic material, Vacuum 162 (2019)
  156--162.

\bibitem{vija2017high}
R.~K. Vijayaraghavan, D.~Chandran, R.~K. Vijayaraghavan, A.~P. McCoy,
  S.~Daniels, P.~J. McNally, Highly enhanced {UV} responsive conductivity and
  blue emission in transparent {CuBr} films: implication for emitter and
  dosimeter applications, Journal of Materials Chemistry C 5~(39) (2017)
  10270--10279.

\bibitem{zhou2023grow}
H.~Zhou, X.~Chen, T.~Zhao, C.~Chen, J.~Han, S.~Pan, J.~Pan, Growth and
  luminescence properties of $\gamma$-{CuBr} single crystals by the bridgman
  method, Cryst.Eng.Comm 25~(11) (2023) 1669--1674.

\bibitem{bies2017adva}
M.~Biesinger, Advanced analysis of copper x-ray photoelectron spectra, Surface
  and Interface Analysis 49 (2017) 1325--1334.

\bibitem{yuan2020engi}
W.~Yuan, Y.~Wu, C.~Fang, X.~Wang, X.~Huang, C.~Li, Engineering nanowire array
  structure of copper oxide/cuprous oxide for enhanced oxygen evolution
  reaction, Journal of The Electrochemical Society 167 (2020) 022508.

\bibitem{xiab2024aself}
B.~Xia, L.~Zhang, D.~Tian, S.~He, N.~Cao, G.~Xie, D.~Zhang, X.~Chu, F.~Zhao,
  {A} self-powered p-{CuBr}/n-{Si} heterojunction photodetector based on vacuum
  thermally evaporated high-quality {CuBr} films, Journal of Materials
  Chemistry C 12 (2024) 1012--1019.

\bibitem{stef2026surf}
S.~A. Irimiciuc, P.~Pokorny, S.~Chertopalov, M.~Vondraek, S.~Cicho, L.~V.
  Lenka, E.~Maresova, R.~Yatskiv, M.~Novonty, J.~Lanok, Surface stability of
  {CuBr} thin films deposited by pulsed laser deposition, Applied Surface
  Science 717 (2026) 164813.

\bibitem{canv2020infl}
N.~Ca, H.~T. Van, P.~Do, L.~Duy~Thanh, P.~Minh~Tan, N.~Truong, V.~Oanh,
  N.~Binh, N.~Hien, Influence of precursor ratio and dopant concentration on
  the structure and optical properties of {Cu}-doped {ZnCdSe}-alloyed quantum
  dots, RSC Advances 10 (2020) 25618--25628.

\bibitem{mfer1996elec}
M.~Ferhat, A.~Zaoui, M.~Certier, J.~Dufour, B.~Khelifa, Electronic structure of
  the copper halides {CuCl}, {CuBr} and {Cul}, Materials Science and
  Engineering: B 39~(2) (1996) 95--100.

\bibitem{vale1998dyna}
J.~Valenta, J.~Moniatte, P.~Gilliot, B.~Honerlage, J.~Grun, R.~Levy, A.~Ekimov,
  Dynamics of excitons in cubr nanocrystals: Spectral-hole burning and
  transient four-wave-mixing measurements, Physical Review B 57 (1998)
  1774--1783.

\bibitem{huon2024eudo}
T.~Huong, N.~Sa, N.~Thuy, H.~Pham, N.~H. Tho, N.~Hin, N.~Ca, {E}u$^{3+}$-doped
  {ZnO} quantum dots: structure, vibration characteristics, optical properties,
  and energy transfer process, Nanoscale Advances 7 (2024) 909--921.

\bibitem{gold1977band}
A.~Goldmann, Band structure and optical properties of tetrahedrally coordinated
  {Cu}- and {Ag}-halides, {P}hysica {S}tatus {S}olidi (b) 81~(1) (1977) 9--47.

\bibitem{serr2002elec}
J.~Serrano, C.~Schweitzer, C.~T. Lin, K.~Reimann, M.~Cardona, D.~Fr\"ohlich,
  Electron-phonon renormalization of the absorption edge of the cuprous
  halides, Physical Reviews B 65 (2002) 125110.

\bibitem{haoy2022band}
H.~Yu, X.~Cai, Y.~Yang, Z.-H. Wang, S.-H. Wei, Band gap anomaly in cuprous
  halides, Computational Materials Science 203 (2022) 111157.

\bibitem{jana2022inve}
B.~Jana, S.~Ghosh, A.~Dutta, A.~V. Baranov, A.~V. Fedorov, A.~Patra,
  Investigation of carrier dynamics of {QD}s using kinetic model and ultrafast
  spectroscopy, Optical Materials: X 13 (2022) 100126.

\end{thebibliography}

%


\newpage
{\normalsize \listoftables}

{\normalsize \listoffigures}




\end{document}